\documentclass{article}

\usepackage[numbers]{natbib} 
    \usepackage[preprint]{neurips_2024}

\usepackage{pdfpages} 
\usepackage[utf8]{inputenc} 
\usepackage[T1]{fontenc}    
\usepackage{hyperref}
\hypersetup{
    colorlinks=true,
    linkcolor=black,
    filecolor=black,      
    urlcolor=black,
    citecolor= black,
    pdftitle={Overleaf Example},
    pdfpagemode=FullScreen,
    }\usepackage{url}            
\usepackage[most]{tcolorbox} 
\usepackage{amsfonts}       
\usepackage{nicefrac}       
\usepackage{microtype}      
\usepackage{xcolor}         
\usepackage{graphicx}
\usepackage{glossaries}
\usepackage{float}
\usepackage[inline]{enumitem}
\makeglossaries
\newtcolorbox[auto counter]{mybox}[2][]{%
colback=blue!5!white, colframe=blue!75!black,floatplacement=p,float,title=Box~\thetcbcounter: #2,#1}
\title{Whether to trust: the ML leap of faith}

\newacronym{json}{JSON}{JavaScript Object Notation}
\newacronym{csv}{CSV}{comma separated values}
\newacronym{cs}{CS}{circadian stimulus}
\newacronym{rgb}{RGB}{red green blue}
\newacronym{ntc}{NTC}{negative temperature coefficient}
\newacronym{uvi}{UVI}{ultraviolet index}
\newacronym{cir}{CIR}{Clutter Image Rating}
\newacronym{pin}{PIN}{personal identification number}

\newacronym{bal}{BAL}{blood alcohol level}

\newacronym{rhr}{RHR}{resting heart rate }
\newacronym{eda}{EDA}{electro-dermal activity}
\newacronym{gp1}{GP}{Gaussian process}
\newacronym{bleu}{BLEU}{BiLingual Evaluation Understudy}
\newacronym{nlp}{NLP}{natural language processing}
\newacronym{ukri}{UKRI}{UK Research and Innovation}
\newacronym{bmi}{BMI}{body mass index}
\newacronym{nist}{NIST}{National Institute for Standards and Technology}
\newacronym{3d}{3D}{declared, demonstrated and deserved trust}
\newacronym{3p}{3P}{predisposing, precipitating, perpetuating factors}
\newacronym{5ht}{5-HT}{Serotonin}
\newacronym{6smt}{6-SMT}{6-sulphatoxymelatonin}
\newacronym{aasm}{AASM}{American Academy of Sleep Medicine}
\newacronym{ab}{A\textbeta}{beta-amyloid}
\newacronym{ach}{ACH}{acetylcholine}
\newacronym{acth}{ACTH}{adrenocorticotropic hormone}
\newacronym{amt6s}{aMT6s}{6-sulphatoxymelatonin, the primary urinary metabolite of melatonin}
\newacronym{aswpd}{ASWPD}{advanced sleep-wake phase disorder}
\newacronym{ad}{AD}{Alzheimer's disease}
\newacronym{asps}{ASPS}{advanced sleep phase syndrome}
\newacronym{ai}{AI }{artificial intelligence}
\newacronym{ahi}{AHI}{Apnea Hypopnea Index }
\newacronym{aie}{AIE}{attention–intention-effort}
\newacronym{ait}{AIT }{artificial intelligence technology}
\newacronym{anova}{ANOVA}{analysis of variance}
\newacronym{api}{API }{Application Programming Interface}
\newacronym{ar}{AR}{augmented reality}
\newacronym{arcs}{ARCS}{Attention Relevance Confidence Satisfaction - a design model}
\newacronym{art}{ART }{accountability, responsibility and transparency}
\newacronym{asq}{ASQ}{Athens Sleep Questionnaire}
\newacronym{bagging}{bagging}{Bootstrap AGGregatING}
\newacronym{bci}{BCI}{brain–computer interface}
\newacronym{bct}{BCT }{behaviour change techniques}
\newacronym{bcw}{BCW}{Behaviour Change Wheel}
\newacronym{bf}{BF}{basal forebrain}
\newacronym{blog}{BLOG}{Bayesian LOGic}
\newacronym{bptt}{BPTT}{back propagation through time}
\newacronym{cahi}{CAHI}{central apnea-central hypopnea index}
\newacronym{cbt}{CBT}{Cognitive Behavioural Therapy}
\newacronym{cobt}{CBT}{core body temperature}
\newacronym{cbtm}{$\acrshort{cbt}_{min}$}{minimum core body temperature}
\newacronym{cbti}{CBT-I}{Cognitive Behavioural Therapy for Insomnia}
\newacronym{cchs}{CCHS}{congenital central hypoventilation syndrome}
\newacronym{cdc}{CDC}{Centers for Disease Control and Prevention}
\newacronym{chd}{CHD}{coronary heart disease}
\newacronym{croni}{CI}{chronic insomnia}
\newacronym{ci}{CI}{confidence interval}
\newacronym{cks}{CKS}{Clinical Knowledge Summaries}
\newacronym{covid}{COVID}{Coronavirus disease}
\newacronym{cns}{CNS}{central nervous system}
\newacronym{crswd}{CRSWD}{circadian rhythm sleep-wake disorder}
\newacronym{cnn}{CNN}{convolutional neural network}
\newacronym{comb}{COM-B}{Capability Opportunity Motivation Behaviour}
\newacronym{cpu}{CPU}{central processing unit}
\newacronym{cra}{CRA }{credit reference agency}
\newacronym{crh}{CRH}{corticotropin-releasing hormone}
\newacronym{crl}{CRL}{Causal reinforcement learning}
\newacronym{crp}{CRP}{C-reactive protein}
\newacronym{csa}{CSA}{central sleep apnea}
\newacronym{csa-csb}{CSA-CSB}{central sleep apnea with Cheyne-Stokes breathing}
\newacronym{csb}{CSB}{Cheyne-Stokes breathing}
\newacronym{csd}{CSD}{Consensus Sleep Diary}
\newacronym{csf}{CSF}{cerebrospinal fluid}
\newacronym{ct}{CT}{circadian time}
\newacronym{cta}{CTA}{chronotype-adjusted}
\newacronym{cvd}{CVD}{cardiovascular disease}
\newacronym{dafti}{DAFTI}{Demonstrated Action Follow-through Index}
\newacronym{db}{dB}{decibels}
\newacronym{dbt}{DBT}{desired bed time}
\newacronym{dis}{DIS}{difficulty initiating sleep}
\newacronym{dirti}{DIRTI}{Demonstrated Intention Relative Trust Index}
\newacronym{difti}{DIFTI}{Demonstrated Intention Follow-through Trust Index}
\newacronym{dlmo}{DLMO}{dim light melatonin onset}
\newacronym{dlmoo}{DLMOff}{dim light melatonin offset}
\newacronym{dms}{DMS}{difficulty maintaining sleep}
\newacronym{dof}{DoF}{Degree of Fragmentation}
\newacronym{drn}{DRN}{dorsal raphe nucleus}
\newacronym{dsm4}{DSM‐IV}{Diagnostic and Statistical Manual of Mental Disorders, 4th edition}
\newacronym{dswpd}{DSWPD}{delayed sleep-wake phase disorder}
\newacronym{dsst}{DSST}{digit symbol substitution task}
\newacronym{dt}{D/T}{due to}
\newacronym{dwa}{DWA}{delta wave amplitude}
\newacronym{ecg}{ECG }{electrocardiogram/ electrocardiography}
\newacronym{ed50}{$ED_{50}$}{effective dose for 50\% suppression}
\newacronym{eeg}{EEG}{electroencephalogram}
\newacronym{ema}{EMA}{early morning awakening}
\newacronym{emg}{EMG}{electromyogram}
\newacronym{edi}{EDI}{equivalent daylight illuminance }
\newacronym{ess}{ESS}{Epworth Sleepiness Scale}
\newacronym{eog}{EOG}{electro-oculogram}
\newacronym{epr}{EPR}{electronic patient record}
\newacronym{eu}{EU}{European Union}
\newacronym{fbcsp}{FBCSP}{filter bank common spatial patterns}
\newacronym{fbm}{FBM}{Fogg Behavioural Model}
\newacronym{fcn}{FCN}{Fully Convolutional Network}
\newacronym{fda}{FDA}{Food and Drug Administration}
\newacronym{fiti}{FITI}{Follow Intention Through index}
\newacronym{gaba}{GABA}{gamma aminobutyric acid}
\newacronym{gad7}{GAD7}{Generalised Anxiety Disorder clinical questionnaire}
\newacronym{gan}{GAN}{Generative Adverserial Network}
\newacronym{gdpr}{GDPR}{General Data Protection Regulation}
\newacronym{gh}{GH}{growth hormone}
\newacronym{ghrh}{GHRH}{growth hormone-releasing hormone}
\newacronym{gp}{GP}{Gaussian process}
\newacronym{gpu}{GPU}{graphics processing unit}
\newacronym{gru}{GRU}{Gated recurrent units}
\newacronym{h}{H}{hours}
\newacronym{hapa}{HAPA}{Health Action Process Approach}
\newacronym{hav}{HAV}{hepatitis A virus}
\newacronym{hbm}{HBM}{Health Belief Model}
\newacronym{hci}{HCI}{Human Computer Interaction}
\newacronym{hh}{HH}{household}
\newacronym{hla}{HLA}{human leukocyte antigen}
\newacronym{hgh}{HGH}{hypnagogic hallucinations}
\newacronym{homair}{HOMA-IR}{estimated insulin resistance}
\newacronym{hpa}{HPA}{hypothalamic–pituitary–adrenal}
\newacronym{hph}{HPH}{ hypnopompic hallucinations}
\newacronym{hrqal}{HR-QAL}{health-related quality of life}
\newacronym{hsat}{HSAT}{home sleep apnea test}
\newacronym{hr}{HR}{hazard ratio}
\newacronym{hr1}{HR}{heart rate}
\newacronym{hrv}{HRV}{heart rate variability}
\newacronym{html}{HTML}{HyperText Markup Language}
\newacronym{hz}{Hz}{hertz}
\newacronym{iapt}{IAPT}{Improving Access to Psychological Therapies}
\newacronym{ibd}{IBD}{inflammatory
bowel diseases }
\newacronym{icc}{ICC}{intraclass correlation coefficient }
\newacronym{icd}{ICD}{International Statistical Classification of Diseases}
\newacronym{icd10}{ICD-10}{ICD, Tenth Revision, Clinical Modification}
\newacronym{icd11}{ICD-11}{ICD, Eleventh Revision, Clinical Modification}
\newacronym{icsd}{ICSD-3}{International Classification of Sleep Disorders 3rd edition}
\newacronym{ih}{IH}{idiopathic hypersomnia}
\newacronym{ieee}{IEEE}{Institute of Electrical and Electronics Engineers}
\newacronym{il1}{IL-1\textbeta}{interleukin-1\textbeta}
\newacronym{il2}{IL-2}{interleukin-2}
\newacronym{il6}{IL-6}{interleukin-6}
\newacronym{igg}{IgG}{immunoglobulin G antibodies}
\newacronym{igm}{IgM}{immunoglobulin M antibodies}
\newacronym{ios}{IOS}{iPhone Operating System}
\newacronym{iot}{IoT}{Internet of Things}
\newacronym{ip}{IP}{intellectual property}
\newacronym{iprgcs}{ipRGCs}{intrinsically photosensitive retinal ganglion cells}
\newacronym{iqr}{IQR}{Inter Quartile Range}
\newacronym{irls}{IRLS}{International Restless Legs Scale}
\newacronym{irt}{IRT}{Image Relief Therapy}
\newacronym{isi}{ISI}{Insomnia Severity Index}
\newacronym{iswrd}{ISWRD}{irregular sleep-wake rhythm disorder}
\newacronym{itt}{ITT}{Intention to Treat}
\newacronym{ittt}{ITTT}{If This Then That}
\newacronym{jld}{JLD}{jetlag disorder}
\newacronym{k}{K}{thousand}
\newacronym{kls}{KLS}{Kleine-Levin syndrome}
\newacronym{knn}{KNN}{K-nearest neighbours}
\newacronym{led}{LED}{Light Emitting Diode}
\newacronym{lhs}{LHS}{left-hand side}
\newacronym{lofm}{LoF}{Leap of Faith}
\newacronym{lstm}{LSTM}{Long short-term memory}
\newacronym{lux}{lux}{luminous flux per unit area (light intensity)}
\newacronym{m}{M}{million}
\newacronym{mcid}{MCID}{minimum clinically important difference}
\newacronym{mctq}{MCTQ}{Munich Chronotype Questionnaire}

\newacronym{md}{MD}{mean difference}
\newacronym{meq}{MEQ}{Morningness‐Eveningness Questionnaire}
\newacronym{met}{MET}{metabolic equivalents}
\newacronym{mets}{MetS}{metabolic syndrome}
\newacronym{mfcc}{MFCCs}{Mel-frequency cepstral coefficients}
\newacronym{mg}{mg}{milligram}
\newacronym{mi}{MI}{myocardial infarction}
\newacronym{midlife}{mid-life}{45-74-year-old}
\newacronym{ml}{ML}{machine learning}
\newacronym{mlux}{mLux}{melanopic \acrshort{lux} – includes wavelength}
\newacronym{mse}{MSE}{mean squared error}
\newacronym{mic}{Mike}{microphone}
\newacronym{mlr}{MLR}{Multivariate linear regression}
\newacronym{mmhg}{mmHg}{millimetres of mercury}
\newacronym{mrna}{mRNA}{messenger ribonucleic acid}
\newacronym{mra}{MRA}{melatonin-receptor agonist}
\newacronym{ms}{ms}{millisecond}
\newacronym{mslt}{MSLT}{multiple sleep latency test}
\newacronym{mturk}{MTurk}{Amazon Mechanical Turk}
\newacronym{n24}{N24SWD}{non-24-hour sleep-wake rhythm disorder}
\newacronym{nan}{NaN}{Not a Number - i.e. no data}
\newacronym{nhs}{NHS }{National Health Service}
\newacronym{nia}{NIA }{NHS Innovation Accelerator}
\newacronym{nihr}{NIHR}{National Institute for Health and Care Research}
\newacronym{nice}{NICE}{National Institute for Health and Care Excellence}
\newacronym{nismd}{NISMD}{Non-invasive sleep-measuring devices}
\newacronym{nk}{NK}{natural killer}
\newacronym{nm}{nm}{nanometer}
\newacronym{nos}{NOS}{SWD not otherwise specified}
\newacronym{nrem}{NREM}{non-rapid eye movement sleep}
\newacronym{nsaid}{NSAID}{non-steroidal anti-inflammatory drug}
\newacronym{nsf}{NSF}{National Sleep Foundation}
\newacronym{ocd}{OCD}{Obsessive Compulsive Disorder}
\newacronym{ocst}{OCST}{Out of Center Sleep Testing}
\newacronym{oecd}{OECD}{Organisation for Economic Co-operation and Development}
\newacronym{or}{OR}{odds ratio}
\newacronym{osa}{OSA}{obstructive sleep apnoea}
\newacronym{pap}{PAP}{positive airway pressure}
\newacronym{pat}{PAT}{peripheral arterial tonometry}
\newacronym{pba}{PBA}{Person Based Approach}
\newacronym{pc}{PC  }{personal computer - includes desktops and laptops}
\newacronym{pca}{PCA}{principal component analysis}
\newacronym{pd}{PD}{primary dysmenorrhea}
\newacronym{pg}{PG}{prostaglandins}
\newacronym{pgml}{pg/ml}{picograms per milliliter}
\newacronym{phenotype}{phenotype}{set of observable characteristics given genotype and environment interaction}
\newacronym{phq9}{PHQ9}{Patient Health Questionnaire for depression}
\newacronym{pi}{PI}{Primary Insomnia}
\newacronym{plmd}{PLMD}{periodic limb movement disorder}
\newacronym{plms}{PLMS}{periodic-limb movements of sleep}
\newacronym{plmw}{PLMW}{periodic-limb movements of wake}
\newacronym{pms}{PMS}{premenstrual syndrome}
\newacronym{pp1}{PP}{Per protocol}
\newacronym{psm}{PSM}{propriospinal myoclonus}
\newacronym{pnt}{PNT}{pneumotachograph/ pneumotachography}
\newacronym{pp}{pp}{percentage point}
\newacronym{ppg}{PPG}{photoplethysmograph/ photoplethysmography}
\newacronym{poa}{POA}{preoptic area of hypothalamus}
\newacronym{poc}{PoC}{Processes of Change}
\newacronym{pods}{PODS}{Personal Online Data Stores}
\newacronym{prc}{PRC}{phase response curve}
\newacronym{prec}{PREC}{Psychology Research Ethics Committee}
\newacronym{aprt}{APRT}{abbreviated progressive muscle relaxation}
\newacronym{psd}{PSD}{Pittsburg Sleep Diary}
\newacronym{psd1}{PSD}{partial sleep deprivation}
\newacronym{psg}{PSG}{polysomnography}
\newacronym{psqi}{PSQI}{Pittsburgh Sleep Quality Index}
\newacronym{ptsd}{PTSD}{Post Traumatic Stress Disorder}
\newacronym{pvt}{PVT}{psychomotor vigilance task}
\newacronym{pox}{POx}{pulse oximetry}
\newacronym{ra}{RA}{rheumatoid arthritis}
\newacronym{ram}{RAM}{RAND/UCLA Appropriateness Method}
\newacronym{rbd}{RBD}{REM sleep behaviour disorder}
\newacronym{rbdss}{RBDSS}{REM sleep behaviour disorder severity scale}
\newacronym{rct}{RCT}{Randomised Control Trial}
\newacronym{rera}{RERA}{respiratory effort- related arousal }
\newacronym{rem}{REM}{rapid eye movement}
\newacronym{rf}{RF}{Random Forest}
\newacronym{rhs}{RHS}{right-hand side}
\newacronym{rip}{RIP}{respiratory inductance plethysmography}
\newacronym{rl}{RL}{Reinforcement Learning}
\newacronym{rls}{RLS}{restless legs syndrome}
\newacronym{rrv}{RRV}{respiratory rate variability}
\newacronym{rnn}{RNN}{Recurrent neural network}
\newacronym{rpfc}{RPFC}{right prefrontal cortex}
\newacronym{roc}{ROC}{Receiver Operating Characteristic}
\newacronym{rr}{RR}{relative risk}
\newacronym{remr}{RR}{remission rate}
\newacronym{rwa}{RWA}{REM without atonia}
\newacronym{sast}{SAST}{serial addition/subtraction task}
\newacronym{sct1}{SCT}{Social Cognitive Theory}
\newacronym{sc}{SC}{Stimulus Control}
\newacronym{sd}{SD}{standard deviation}
\newacronym{sdb}{SDB}{sleep-disordered breathing}
\newacronym{sdk}{SDK}{Software Development Kit}
\newacronym{se}{SE}{Sleep Efficiency}
\newacronym{sh}{SH}{sleep hygiene}
\newacronym{shap}{SHAP}{SHapley Additive exPlanations}
\newacronym{sit}{SIT}{Suggested Immobilisation Test}
\newacronym{soc}{SoC}{Stages of Change}
\newacronym{sol}{SOL}{Sleep onset latency}
\newacronym{soremps}{SOREMPs}{sleep onset REM periods}
\newacronym{srh}{SRH}{sleep-related hypoventilation}
\newacronym{srhd}{SRHD}{sleep-related hypoventilation disorders}
\newacronym{srlc}{SRLC}{sleep-related leg cramps}
\newacronym{srmd}{SRMD}{sleep-related movement disorder}
\newacronym{srrmd}{SRRMD}{sleep-related rhythmic movement disorder}
\newacronym{srt}{SRT}{Sleep Restriction Therapy}
\newacronym{srva}{SRVAs}{sleep-related vehicle accidents}
\newacronym{ssm}{SSM}{sleep state misperception}
\newacronym{sss}{SSS}{Stanford Sleepiness Scale}
\newacronym{stem}{STEM}{Science, Technology, Engineering and Mathematics}
\newacronym{swa}{SWA}{slow-wave activity}
\newacronym{swd}{SWD}{shift-work disorder}
\newacronym{swpd}{SWPD}{sleep-wake phase disorder}
\newacronym{sws}{SWS}{Slow-Wave Sleep}
\newacronym{tbc}{TBC}{To be confirmed}
\newacronym{tbr}{TBR}{time in bed regularisation}
\newacronym{tdi}{TDI}{Townsend Deprivation Index}
\newacronym{tib}{TIB}{Time in bed}
\newacronym{tpb}{TPB}{Theory of Planned Behaviour}
\newacronym{tra}{TRA}{Theory of Reasoned Action}
\newacronym{tsd}{TSD}{total sleep deprivation}
\newacronym{tsh}{TSH}{thyroid stimulating hormone}
\newacronym{tst}{TST}{Total Sleep Time}
\newacronym{ttm}{TTM}{Transtheoretical Model}
\newacronym{twt}{TWT}{total wake time}
\newacronym{ubm}{UBM}{universal background model}
\newacronym{uv}{UV}{ultraviolet}
\newacronym{ux}{UX}{user experience}
\newacronym{vlpo}{VLPO}{ventrolateral preoptic area}
\newacronym{vms}{VMS}{vasomotor symptoms}
\newacronym{vpsg}{vPSG}{video polysomnography}
\newacronym{waso}{WASO}{wake after sleep onset}
\newacronym{doti}{DOTI}{Deserving Of Trust index}
\newacronym{wbc}{WBC}{white blood cells}
\newacronym{who}{WHO}{World Health Organisation}
\newacronym{woa}{WOA}{Weight on Advice}
\newacronym{xai}{XAI}{Explainable \acrshort{ai}}

\author{%
  Tory Frame \thanks{We would like to thank all study participants for investing their valuable time in developing and testing our sleep-improvement system. Furthermore, we wish to thank Deborah Morgan, Madalin Facino, and Eoin Cremen for their advice on earlier versions of this paper and Tom Donnelly for help developing the Sleep Angel device.} \\
  Department of Computer Science\\  
  University of Bath\\
  \texttt{tvhf20@bath.ac.uk} \\
  \And
    George Stothart \\ Department of Psychology 
 \\
  University of Bath \\
  \texttt{gs744@bath.ac.uk} \\
  \AND
  Elizabeth Coulthard \\  
  University of Bristol Medical School \\ North Bristol NHS Trust \\
  \texttt{Elizabeth.Coulthard@bristol.ac.uk} \\
    \And
    Julian Padget \\ Department of Computer Science\\ University of Bath
 \\
  \texttt{masjap@bath.ac.uk} \\
}

\begin{document}

\maketitle

\begin{abstract}
Human trust is a prerequisite to trustworthy \acrshort{ai} adoption, yet trust remains poorly understood. Trust is often described as an attitude, but attitudes cannot be reliably measured or managed. 
Additionally, humans frequently conflate trust in an AI system, its machine learning (\acrshort{ml}) technology, and its other component parts. Without fully understanding the `leap of faith' involved in trusting \acrshort{ml}, users cannot develop intrinsic trust in these systems.
A common approach to building trust is to explain a \acrshort{ml} model's reasoning process. However, such explanations often fail to resonate with non-experts due to the inherent complexity of \acrshort{ml} systems and explanations are disconnected from users’ own (unarticulated) mental models.
This work puts forward an innovative way of directly building intrinsic trust in \acrshort{ml}, by discerning and measuring the \acrfull{lofm} taken when a user decides to rely on \acrshort{ml}. The \acrshort{lofm} matrix captures the alignment between an \acrshort{ml} model and a human expert's mental model. This match is rigorously and practically identified by feeding the user's data and objective function into both an \acrshort{ml} agent and an expert-validated rules-based agent---a verified point of reference that can be tested a priori against a user's own mental model. This represents a new class of neuro-symbolic architecture. The \acrshort{lofm} matrix visually contrasts the different agent outputs, revealing to the user the distance that constitutes the leap of faith between the rules-based and \acrshort{ml} agents. 
For the first time, we propose trust metrics that evaluate whether users demonstrate trust through their actions rather than self-reported intent and whether such trust is deserved based on outcomes. 
The significance of the contribution is that it enables empirical assessment and management of \acrshort{ml} trust drivers, to support trustworthy \acrshort{ml} adoption.  The approach is illustrated through a long-term high-stakes field study: a 3-month pilot of a multi-agent sleep-improvement system.

\end{abstract}

\section{Introduction}
\label{intro}
Lack of trust is often cited as a barrier to \acrfull{ai} adoption, especially for \acrfull{ml}.
Trust has always been a critical enabler of new technology adoption: people tend to rely on automation they trust, and shun automation they distrust, in the real world \citep{zuboff1988age} and the laboratory \citep{muir1996trust}.  However, \acrshort{ml} faces some unique challenges, which are discussed in the context of the psychology, management, and computer-science literatures. 


Trust is commonly understood as a unitary concept but, in practice, it is complex. Lee and Sees' human-centred definition is widely accepted: \textit{the attitude that the technology will help us achieve our goals \citep{lee2004trust} in risky circumstances} \citep{jacovi2021formalizing}, i.e. in an uncertain and vulnerable situation \citep{mayer1995integrative}.  
Initial trust will vary for a technology because we each have a different general tendency to trust due to factors like our culture or personality (`dispositional' trust); and a different ability to deal with the situation and the technology (`situational' trust) \citep{hoff2015trust}.  Once we experience a specific technology, we can develop `learned' trust.  
Most of the discussion of trust in the literature relates to cognitive or rational drivers of which the user is aware.
Lee and Sees' \textit{`performance'/ `process'/ `purpose'\/} framework \citep{lee2004trust} encapsulates most of these conscious drivers. They argue potential for trust in technology is highest when: the trustee perceives `performance' as strong; the trustee understands how the `process' operates \citep{sheridan1992telerobotics}; and the trustee believes its `purpose' is aligned to their own goals. This framework is now used to reflect on the trust challenges faced by \acrshort{ml}.

\acrshort{ml} should, in theory, garner higher trust when \textit{performance} is better than alternatives. For example, a deep-learning model outperformed a team of 6 radiologists by a statistically-significant 11.5\% in a 27,367-woman UK and US breast-cancer study, measured by relative area under the \acrshort{roc} curve \citep{mckinney2020international}.
However, if these technologists tested model output against what they expected given the inputs, they would disagree on outputs where the model outperformed them: false positives and negatives were 3.9-15.1\% (UK-US) higher than the model. 

\acrfull{xai} research motivated by a desire to improve our understanding of the `\textit{process}'
of how \acrshort{ml} models' algorithms work \citep{wortham2020transparency}.
If we understand a reasoning process and it matches our prior knowledge of sensible reasoning, we can directly build intrinsic trust; if not, only extrinsic trust is possible: we need external reassurance from an expert or an evaluation \citep{jacovi2021formalizing}.  \acrshort{xai}'s efforts will only increase intrinsic trust if humans can explicitly compare the \acrshort{ml} model's reasoning to their own mental model \citep{jacovi2021formalizing} and their internal representation of how the world works \citep{craik1967nature}. 
This is problematic because a given explanation does not always fit recipient mental models. 
Because \acrshort{ml} works by finding the best fit for mathematical models with large data sets \citep{prince2023understanding}, it can be challenging for people to follow its reasoning process because it is so different from their own. 
Tools like feature importance or Shapley causal quasi-values \citep{frye2020shapley} reflect the factors that contributed to an \acrshort{ml} model's conclusion a posteriori \citep{holzinger2019causability}, but not its reasoning. 
For example, clinicians can think in terms of an evidence-based mechanism of action \citep{bienefeld2023solving}. Each human-in-the-loop may have a unique a priori mental model, which we only start to discover when \acrshort{ml} outputs do not align with their implicit expectations.
When \acrshort{ml} outperforms humans, there should always be a mismatch between the model's reasoning and expert priors, because new relationships are identified.  Trust can thus, at best, be extrinsic, relying on external assurances, so even domain experts require a leap of faith to act: from what they expect to what the model outputs. 
However, understanding a \acrshort{ml} model's \textit{`process'} requires comprehension of more than algorithmic reasoning: data is also critical and could be understandable, even if the data set is large.  Zytek et al. propose that an interpretable text feature space is: phrased in everyday language avoiding codes (‘readable’); worded so concepts are quick and simple to absorb (‘human-worded’); and reliant on real-world concepts (‘understandable’) \citep{zytek2022need}. 

If a technology's \textit{purpose} is clearly communicated, it is generally trusted to do what it was built to do \citep{lee2004trust}, although there could be concerns about its performance. 
Because humans do not directly control how an  \acrshort{ml} algorithm reaches a conclusion, \acrshort{ml} represents a fundamental shift in agency between humans and technology \citep{murray2021humans}; especially if workflows are automated, and made up of agents that perceive and act \citep{russell2016artificial}. If intent is not communicated, lack of agency could exacerbate concerns about purpose alignment, especially if a system has multiple agents: it will be harder to disentangle how its purpose relates to an individual's own goals, or to trust it will make appropriate trade-offs, given other priorities.

Trust can also be emotional \citep{mcallister1995affect} and/ or have automatic or unconscious drivers, like impulses from learned associations and innate biases \citep{strack2004reflective}. Jacovi et al. suggest if we base our trust on such factors that are unrelated to model trustworthiness, our trust is undeserved \citep{jacovi2021formalizing}. However, a trustworthy system could ethically employ such techniques e.g., if unconscious drivers preclude acceptance.
Glikson et al. explore some relevant tools: 
embodiment and immediacy behaviours \citep{glikson2020human}.
Humans find embodied \acrshort{ai}~-- e.g. physical robots \citep{lee2006physically}~-- more tangible and thus easier to trust \citep{glikson2020human}. 
It is harder to tell whether a system contains \acrshort{ai} if the \acrshort{ai} is embedded, or is part of a larger system: the role of \acrshort{ai} is less tangible and therefore harder to perceive, even if its presence is disclosed.  
Agents are digitally, not physically, embodied so sit between the two. 
Immediacy behaviours, such as personalisation, increase perceptions of interpersonal closeness \citep{mehrabian1967attitudes} so can improve \textit{perceived} performance and purpose alignment. 
 
Following the above brief presentation of aspects of trust, we propose a practical way of measuring and managing trust in \acrshort{ml}, that discerns the leap of faith, and measures both demonstrated and deserved trust, enabling trust to be managed over time. We illustrate this using data from a long-term high-stakes field study: a 3-month pilot for a sleep-improvement system with \acrshort{ai} agents. We first review related work in Section~\ref{sec:related}, then briefly describe our study methods in Section \ref{methods} to provide the context for the novel architecture we propose in Section \ref{architecture}. Section \ref{discussion} evaluates the contribution of the architecture and its limitations. 

\section{Related work}
\label{sec:related}

\begin{figure}[!b]
     \centering
     \includegraphics[width=0.9\textwidth, keepaspectratio]{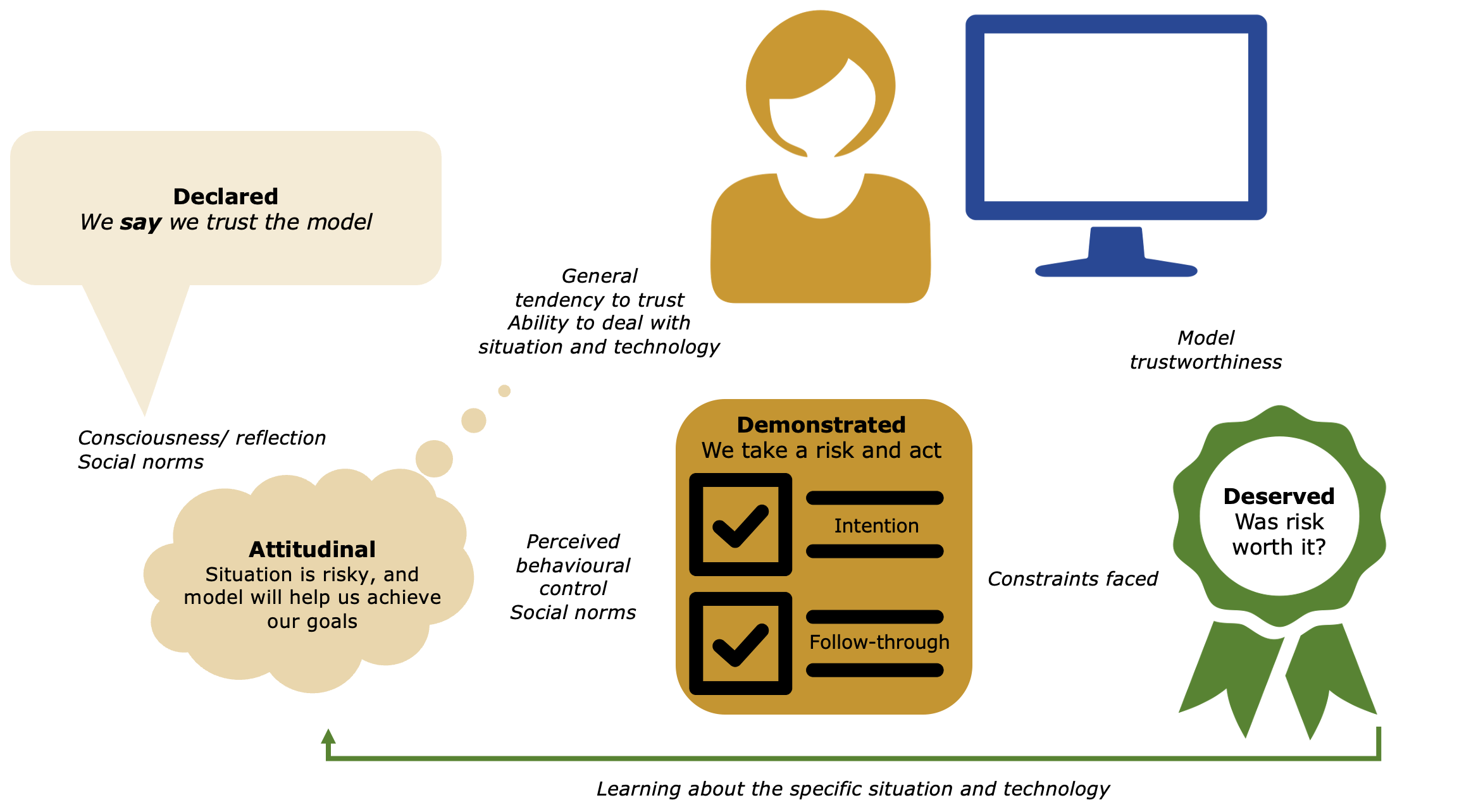} 
     
     \caption{Trust and trustworthiness definitions. All human-centred concepts are shown in yellow; technology ones in blue; combined in green. Icons by Angriawan Ditya Zulkamain, Arif Hariyanto, Karen Tyler and PEBIAN from thenounproject.com.}
\label{trust_framework}
\end{figure}

The literature lacks a hierarchy of tools to support trustworthy \acrshort{ml} adoption. Our inability to directly measure attitudinal trust means that we cannot objectively assess what drives it. However, what matters most for \acrshort{ai} adoption is whether people rely on \acrshort{ai} recommendations, and whether this trust was deserved.  These concepts are depicted in Figure \ref{trust_framework} and will now be explored. We conclude that new metrics are required to measure demonstrated and deserved trust.

%

\subsubsection*{Declared trust}
What is routinely measured, and labeled as attitudinal trust, is \textit{declared trust or what people say they trust} \citep{buccinca2020proxy}. 
It is wrong to label this attitudinal trust, because it only reveals what the respondent is conscious of and is prepared to disclose, given social norms, so it will miss automatic drivers of which the user is unaware, like biases. 
Declared trust is measured subjectively in interviews and surveys using scales that are different and researcher-defined, so it is hard to learn across studies \citep{hoffman2018metrics} and key trust drivers may be missed. 
Declared trust in robots starts higher for the same level of model intelligence and deteriorates over time as performance is evaluated; whereas for embedded \acrshort{ai} trust starts lower and builds over time \citep{glikson2020human}. 
Underlying this insight is an issue with most trust measurement, especially for embedded \acrshort{ml}: it is trust in the system that is measured, not trust in the \acrshort{ml} algorithm as distinct from data or objective function.

\subsubsection*{Demonstrated trust}
\textit{Demonstrated trust is an action or a behaviour}: we take a risk and act, displaying `reliance' \citep{lee2004trust}. 
Demonstrated trust reflects both conscious and automatic drivers but it is not the same as attitudinal trust ~-- e.g. a user may trust but not act because they lack behavioural control, or fear being judged by society; or they might set an intention to act, but not fulfill this intention (or `follow through') because of constraints encountered \citep{ajzen1985intentions}. 
Demonstrated trust can be objectively measured. For example, percentage of recommendations relied upon \citep{lai2019human} or \acrlong{woa}, which measures the extent to which participants adjust their decisions based on the advice provided by an algorithm or a human \citep{logg2019algorithm}. However, these metrics only measure intention and fail to reflect action follow-through, which is critical for adoption, especially if there are multiple constraints to overcome. 
The relationship between declared trust and follow-through to action may be strongest when there is high cognitive complexity ~-- e.g. complex automation and novel situations \citep{lee2004trust}. 
Reliance can also be stronger when an individual has high decisional freedom as to subsequent action and is able to compare the technology's recommendation to the human alternative \citep{hoff2015trust}.

\subsubsection*{Deserved trust}
Warranted, or \textit{deserved trust, is when the risk was worth it because the technology itself was trustworthy} for the use-case \citep{lee2004trust}. Trustworthiness is a property of technology, not a human, so is independent of trust \citep{lee2004trust}: humans can distrust and therefore fail to adopt trustworthy \acrshort{ai}; humans can trust and adopt untrustworthy \acrshort{ai}. Trustworthiness is ideally calibrated to trust so technology is not misused, disused or abused, reducing safety risks \citep{parasuraman1997humans}.

There is substantial discussion on measuring model trustworthiness ~-- e.g. the EU criteria include human agency and oversight; transparency; diversity, non-discrimination, fairness; 
 as well as prevention of harm: technical robustness and safety; privacy and data governance; accountability; and societal and environmental well-being \citep{eu_AI_trustworthy_AI}. 
 
However, the literature on measuring deserved trust is limited -- i.e. was the individual right to act upon their trust in the model? 
Jacovi et al. suggest one approach is to manipulate model trustworthiness so it is no longer calibrated with attitudinal trust, and see how much attitudinal trust changes \citep{jacovi2021formalizing}. This is a conceptual approach that is challenging to put into practice in the real world in a verifiable, controlled, reproducible, comparable manner. Furthermore, it suggests trust changes in a linear manner which may not be true (e.g. it could be step-wise). 
This approach also fails to assess what we really want to know: did trusting the recommendation deliver better outcomes? Demonstrated trust is key, not attitudinal, because a recommendation needs to be acted on for impact to be assessed. The individual's experience in a specific use-case will provide learning as to whether trust was deserved or not, which should then translate back into attitudinal trust and, over time, declared and demonstrated trust. 
While a model might be technically robust, it might not have taken into account factors that would influence action -- e.g. an individual's perception of behavioural control or situation-specific constraints. 
There has been very limited exploration of outcomes in the literature. When mortality outcomes for clinician decisions in treating sepsis in intensive care were assessed , 90-day mortality was found to be lowest when the doctors' dosage was in line with an \acrshort{rl} model's recommendations \citep{komorowski2018artificial}, indicating this model was deserving of trust. The issue with this experiment is that it is that the researchers did not have the option to follow a \acrshort{ml} recommendation rather than their own, so the comparison was done ex-post.  The value of following the \acrshort{ml} recommendation is therefore not directly evaluated -- it is modeled, using a high-confidence off-policy
evaluation method. Other examples in this space simulate clinical environments \citep{schmidgall2024agentclinic}, rather than providing decision makers with decision-making tools in the real world and assessing the value when decision makers follow an \acrshort{ml} recommendation. 

\section{Methods}
\label{methods}


The sleep-improvement study was high stakes because it impacted health \citep{sambasivan2021everyone}, and the scale of required changes were substantial. 
Most participants were short sleepers~-- i.e. their \acrfull{tst} was less than 7 hours a night ~-- 
which is associated with poor health outcomes, e.g. 1.2-1.4x higher Alzheimer's disease risk \citep{sabia2021association}; lower immune response, so short sleepers are 5x more susceptible to infections like the common cold \citep{prather2015behaviorally}; slower reaction times \citep{van2003cumulative}, similar to alcohol intoxication \citep{williamson2000moderate} leading to accidents -- e.g. 1.8x higher in the workplace \citep{aakerstedt2002prospective}. 
The main treatment for those who are not chronic insomniacs is ‘sleep hygiene’ (\acrshort{sh}): up to 16 behavioural and environmental suggestions (e.g. avoid light in the evening). While \acrshort{sh} interventions have a medium effect in healthy-adult sleep trials \citep{murawski2018systematic}, research has been subjective and failed to take into account individual sensitivity, even though it varies greatly: e.g. one person’s sensitivity to evening light can be 40x that of another person \citep{phillips2019high}. 
\acrfull{sws} clears beta amyloid plaques, linked to Alzheimer's disease, from the brain \citep{fultz2019coupled}. There has been limited research into the impact of \acrshort{sh} practices on healthy-adult \acrshort{sws}. 
The situation was thus uncertain, so participants took a risk when they followed a recommendation.

\begin{figure}[!t]
     \centering
     \includegraphics[width=0.9\textwidth]{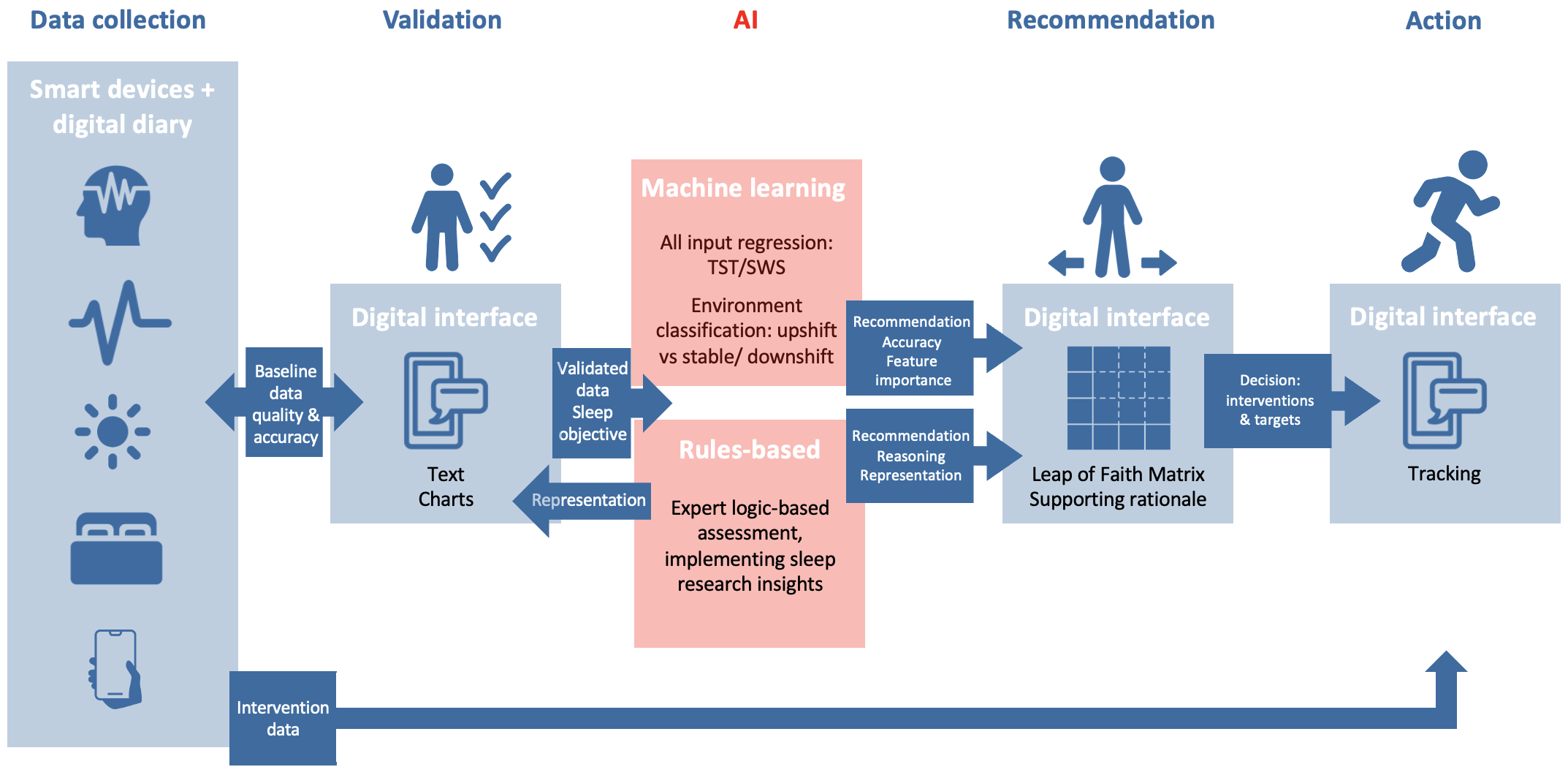} 
     \caption{Socio-technical sleep-improvement system with parallel neuro-symbolic AI. 
}
\label{model}
\end{figure}

The sleep-improvement system was built in collaboration with the target audience: 35 healthy adults -- i.e. no chronic insomnia, \acrlong{osa}, severe anxiety or depression~-- and tested in a pilot with ten healthy 40-55-year-olds. 
Figure \ref{model} illustrates our multi-agent behaviour-support socio-technical system \citep{trist1981evolution}, i.e. it includes both machines and humans. Smart devices collect data; \acrshort{ai} agents process data; a user interface delivers a user's recommendations and tracking through their smartphone. A human is responsible for data collection and validation; objective function specification; and recommendation selection and subsequent implementation. 
The \acrshort{ai} is neuro-symbolic, combining \acrshort{ml} models with a rules-based model: Neuro$\parallel$Symbolic, extending the notation of Kautz \cite{kautz2022third}, signifying that the models or agents work in parallel, producing two recommendations for the user. We believe that this parallel design represents a new form of neuro-symbolics, in addition to the five technical designs identified by Kautz. 

Our models were designed to be trustworthy using the EU criteria \citep{eu_AI_trustworthy_AI}: human agency and oversight, and transparency are central our socio-technical-system design. We implemented the other principles as follows:
\begin{enumerate*}
\item Models were run on each individual's own data. \item Participants were informed of model technical robustness and safety: we informed them that rules-based model's recommendations were sleep-expert approved and of their own \acrshort{ml} model accuracy: 89\% for \acrshort{tst} (81-92\% range, 3.5\% \acrshort{sd}); and 81\% for \acrshort{sws}  (75-86\%, 4.5\% \acrshort{sd}).  
\item Privacy and data standards were specified by the University Ethics Committee (PREC reference number 22 003, 5th September 2022). These standards were a key selection criteria for smart devices, requiring the team to create their own environmental monitor because no third-party solution was identified that delivered sufficient granularity and met these standards. \item The University Ethics Committee holds the researcher accountable for model trustworthiness and upholding standards, and each participant was accountable for their own data, objective function, and choices. 
\item
The key societal challenge was ensuring participant sleep anxiety did not increase as a consequence of the study. Sleep anxiety was measured before and after the study, and it (statistically insignificantly) declined. Severe anxiety or depression were screened out, through \acrshort{gad7} and \acrshort{phq9} questionnaires. Sleep end points were continually monitored, with an 85\% floor on \acrlong{se}, the percentage of time participants were asleep while in bed. 
\end{enumerate*}

\section{Proposed architecture}
\label{architecture}

Human systems of trust, in high-stakes environments like medicine, rely on trusted colleagues pressure-testing conclusions based on validated inputs and standard practices, given desired patient outcomes \citep{bion2023peer}. 
Our participants validated input data and set the agent's objectives; experts tested whether our rules-based \acrshort{ai} model reflected their mental model: they interrogated the logic with our validated, to test whether their standard practices had been correctly implemented.
This model-architecture functional compartmentalisation therefore enables user oversight over data and objective function, increasing their agency. Our expert-validated rules-based \acrshort{ai} model creates a verified reference agent that can be tested a priori by \acrshort{ml} model users. Our \acrshort{lofm} matrix visually contrasts this reference agent to our \acrshort{ml} agent’s recommendation, using the same data and objective function. Areas of agreement can be readily trusted and what was unique to the \acrshort{ml} agent can be identified and addressed. 
We propose three simple objective metrics to measure demonstrated trust, distinguishing intention-setting and follow-through, and whether trust was deserved. 

\subsection{Enabling human agency and oversight} 
\label{model_functional}

Model-architecture functional compartmentalisation is used to separate data and objective function from model reasoning. Each user was responsible for pre-validating their own data.  This exercise uses text and data visualisation techniques, extending the text-driven interpretable-feature-space approach of Zytek et al. \citep{zytek2022need} to include target variables and adds data visualisation, to make the data more accessible to users who benefit cognitively from visual processing. 

\begin{figure}[ht]
     \centering
     \includegraphics[width=\textwidth, keepaspectratio]{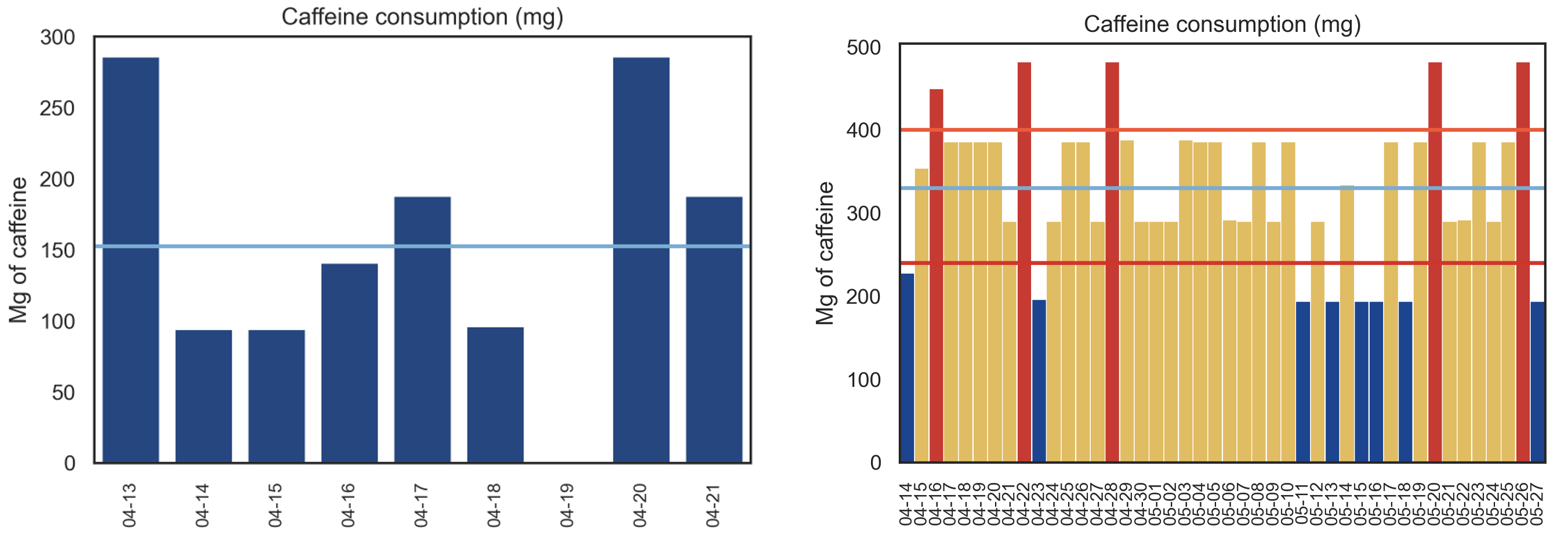} 
     
     \caption{Caffeine data validation example: upfront data-quality trial chart on the left; baseline chart on the right. The rules embedded in the rules-based model were used to colour-code the latter – in this example, blue for <230mg; red for >400mg; yellow for in-between.  
}
\label{validation}

 \end{figure}

Validation started with data quality~-- percentage of each day with complete data for each device~-- which a data-quality agent reported daily and weekly against a pre-briefed target (70\%). Participants progressed to baseline, the next phase of the study, if their one-week trial data was over target. 
Participants then assessed whether the data collected during their trial week reflected their perception.  Each feature input was provided in a text and visual format that had been reviewed by sleep experts and by 35 participants to ensure each was readable, human-worded and understandable, e.g.: `You consume 152mg of caffeine a day (0-286mg range)', and the chart on the left-hand side of Figure \ref{validation}.   

  \begin{figure}[ht]
     \centering
     \includegraphics[width=\textwidth, keepaspectratio]{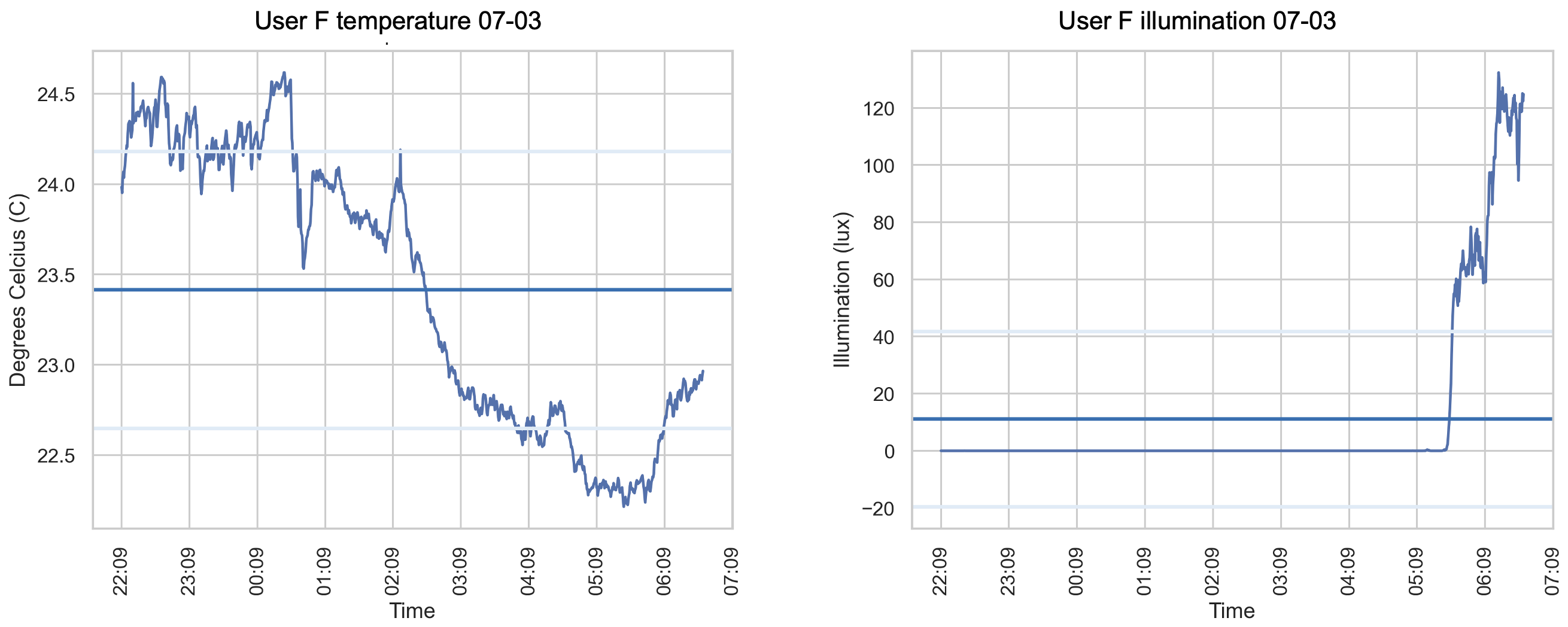} 
     
     \caption{Unfamiliar-data example recorded while the participant was asleep: bedside temperature on the left; bedside illumination on the right. The nightly average is shown as a horizontal mid-blue line.
}
\label{validation_unfamiliar}

 \end{figure}
 
Participants were educated on how to read each chart, what the units meant, and how they were calculated. Participants were encouraged to challenge data accuracy. Some data required considerable interrogation and education because the data and/ or the units were unfamiliar, especially if they were asleep when the data was recorded (e.g. temperature and illumination, as shown in Figure \ref{validation_unfamiliar}).
Participants trialed different behaviours for unfamiliar items to see how the charts responded. Some changed how an input was recorded if the issue was a diary-input mistake. These follow-ups continued until participants were comfortable their data was accurate.
Participants then reviewed at least 1 month of complete, quality days of baseline data, to validate that the target variables and feature inputs were accurate. This included text and charts with sleep-expert-validated logic from the rules-based agent, e.g.: `You had 330mg of caffeine/ day (blue line). You were over 230mg, the amount which impacts the average person’s sleep, on 82\% of days (yellow); over NHS’s 400mg recommended max on 11\% of days (red).' and the chart on the right-hand side of Figure \ref{validation}.

Each participant had unique sleeping challenges (e.g. waking in the night or too early or having trouble sleeping at night) and objectives: some already achieved well over 7 hours' sleep and just wanted to improve \acrshort{sws}; some only got 6 hours of sleep and wanted to focus on \acrshort{tst}.  Sleep objectives were set after the 1-week trial and confirmed once users validated their baseline data and were educated on what was realistic. They thus set the agent's objective function, increasing their agency. 
Participants were given the opportunity to remove any baseline data that they did not believe was accurate, but none did. This is an example of demonstrating trust in the data component of an \acrshort{ml} model, and of a non-expert being able to build intrinsic trust in part of a complex \acrshort{ml} model, because they are able to understand the data inputs and objective function, albeit with support. Data validation and custom question responses, is an example of personalisation, an immediacy behaviour, also increasing agent trustworthiness by ensuring correct inputs.

\subsection{Discerning the ML leap of faith}
\label{discern}

\begin{figure}[!t]
     \centering
     \includegraphics[width=0.7\textwidth, keepaspectratio]{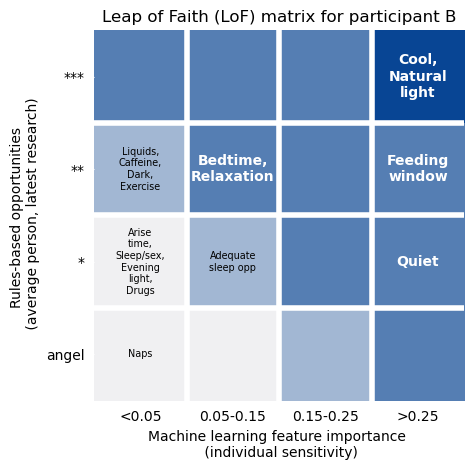} 
     
     \caption{\acrshort{lofm} matrix for a pilot participant.
}
\label{lofm}

 \end{figure}

The recommendation was visually depicted as an intervention `menu' using a \acrshort{lofm} matrix which, as shown in Figure~\ref{lofm}, discerns the leap of faith required to trust the ML agents. 
By using a separate a priori expert-validated rules-based AI agent, we contrast the \acrshort{ml} agents' prioritisation of which \acrshort{sh} changes would make the most difference to the user's sleep objective, with what an expert would conclude from the same inputs. The user is able to see where the agents agree, and where they do not.
Distance from best practice, assessed by the rules-based agent, is used to prioritise opportunities on the vertical axis; \acrshort{ml}-agent feature importance is used to prioritise them on the horizontal. 

The further away an individual was on best practice, the more impact changing that factor could have on the individual's sleep.
The rules-based agent assigned 1 of 3 levels of opportunity to each intervention, or a 4th category: `angel', meaning they had achieved best practice. The 35 design-phase participants preferred 3 levels of opportunity, associating them with a `good, better, best' paradigm, and appreciated the self-efficacy boost of knowing they were already achieving best practice. 
Four demarcations were also used to categorise \acrshort{ml} opportunities using feature importance, measured through average gain across all splits where a feature was used. The higher a feature’s importance, the higher an individual’s sensitivity to that feature. Feature-importance thresholds were set so each individual had at least 4 opportunities in the last 2 columns of the \acrshort{lofm} matrix, equivalent to 2- and 3-star rules-based opportunities.  The \acrshort{ml} agent was more discriminating, e.g. double the number of 3-star opportunities in Figure \ref{lofm}, and 9 `angels', compared to 1 `angel' in the rules-based model. 
Participants were informed that the vertical axis was the expert conclusion based on their data and objective, and the horizontal axis was the \acrshort{ml} assessment of their individual sensitivity using the same inputs. 
The supporting rationale, combining the individual’s validated data with sleep-research insights, was displayed in the user interface, using the rules-based model logic.

Participants were instructed to select interventions they: understood; were able to control and motivated to change \citep{wallston1987perceived}. They were required to give themselves an `adequate opportunity' for 7 hours of sleep, based on \acrshort{se} (\% of time asleep while in bed). They were given decisional freedom to choose up to 3 other interventions on the \acrshort{lofm} matrix. 
Once they chose their interventions, they set specific targets (e.g. get up between 6:45 and 7:15 every day, including weekends).
The user-interface tracker provided daily feedback over the course of 1 month of intervention on how participants were doing against these actions. The user-interface tracker was a simple dashboard accompanied by an automated text message each day. 
The user could investigate each metric, using the same data visualisation as they had used to validate their data (now colour-coded relative to their target).

     
%

\subsection{Measuring the ML leap of faith}

\begin{figure}[!t]
     \centering
     \includegraphics[width=0.8\textwidth, keepaspectratio]{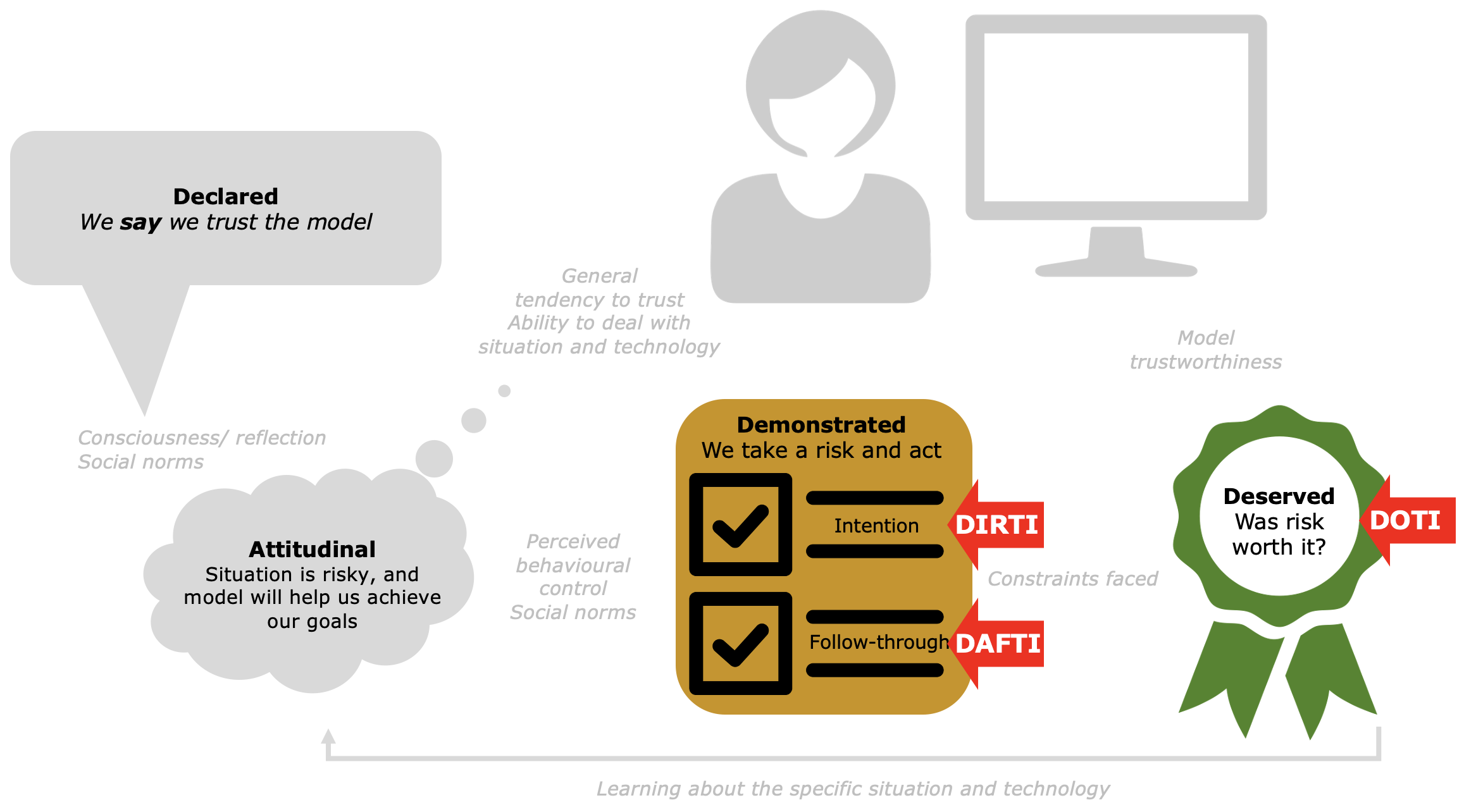} 
     
     \caption{\acrshort{dirti}, \acrshort{dafti} and \acrshort{doti} target demonstrated trust (intention-setting; action-follow); and deserved trust. Icons by Angriawan Ditya Zulkamain, Arif Hariyanto, Karen Tyler and PEBIAN from thenounproject.com CC BY 3.0.}
\label{metrics_summ}

 \end{figure}

 Intervention choice, compliance, and sleep outcomes were recorded, enabling three relative-trust metrics to be assessed: \acrshort{dirti}, \acrshort{dafti}, and \acrshort{doti}. We illustrate each using pilot data. See Figure~\ref{metrics_summ} for how these metrics relate to the types of trust discussed in the literature, and Box~\ref{box:trust metric formulae} for formulae. 
 
\begin{mybox}[label={box:trust metric formulae}]{Trust metric formulae}
\label{dirti_formula}
\[
\textbf{\text{DIRTI}} = \frac{\overline{\text{ML\_raw}}}{\overline{\text{Rules\_raw}}}
\]
\[
\overline{\text{ML\_raw}} = \frac{\sum_{i=1}^n \text{ML\_raw}_i}{n}
\]
\[
\overline{\text{Rules\_raw}} = \frac{\sum_{i=1}^n \text{Rules\_raw}_i}{n}
\]
\[
\textbf{\text{DAFTI}} = \frac{\overline{\text{ML\_adjusted}}}{\overline{\text{Rules\_adjusted}}}
\]
\[
\overline{\text{ML\_adjusted}} = \frac{\sum_{i=1}^n \left( \text{ML\_raw}_i \times \text{Baseline Indicator}_i \right)}{n}
\]
\[
\overline{\text{Rules\_adjusted}} = \frac{\sum_{i=1}^n \left( \text{Rules\_raw}_i \times \text{Baseline Indicator}_i \right)}{n}
\]
\[
\text{Baseline Indicator}_i = 
\begin{cases} 
1 & \text{if Performance}_i > \text{Baseline}_i, \\
0 & \text{otherwise.}
\end{cases}
\]
where: 
\begin{itemize}[nosep]
    \item \(\text{ML\_raw}_i\): \(i\)-th selected intervention raw ML priority score.
    \item \(\text{Rules\_raw}_i\): \(i\)-th selected intervention raw rules-based priority score.
    \item \(n\):  total number of interventions chosen.
    \item \(\sum_{i=1}^n\): summation over all \(n\) individual adjusted scores.
    \item \(\text{Baseline Indicator}_i\): binary variable that equals 1 if performance improved in the last 7 days of intervention against baseline; 0 otherwise.
    \item \(\text{Performance}_i\): last 7-day intervention period average for the \(i\)-th action.
    \item \(\text{Baseline}_i\): baseline average for the \(i\)-th action.
\end{itemize}
\[
\textbf{\text{DOTI}} = \frac{R_{\text{ML\_x}, \text{Outcome}}}{R_{\text{Rules\_x}, \text{Outcome}}}
\]
\[
R_{\text{ML\_x}, \text{Outcome}} = 
\frac{\sum_{i=1}^n \left( \text{ML\_x}_i - \overline{\text{ML\_x}} \right) \left( \text{Outcome}_i - \overline{\text{Outcome}} \right)}
{\sqrt{\sum_{i=1}^n \left( \text{ML\_x}_i - \overline{\text{ML\_x}} \right)^2} \sqrt{\sum_{i=1}^n \left( \text{Outcome}_i - \overline{\text{Outcome}} \right)^2}}
\]
\[
R_{\text{Rules\_x}, \text{Outcome}} = 
\frac{\sum_{i=1}^n \left( \text{Rules\_x}_i - \overline{\text{Rules\_x}} \right) \left( \text{Outcome}_i - \overline{\text{Outcome}} \right)}
{\sqrt{\sum_{i=1}^n \left( \text{Rules\_x}_i - \overline{\text{Rules\_x}} \right)^2} \sqrt{\sum_{i=1}^n \left( \text{Outcome}_i - \overline{\text{Outcome}} \right)^2}}
\]
where: 
\begin{itemize}[nosep]
    \item $R_{\text{ML\_x}, \text{Outcome}}$: correlation between ML priority score \& percentage change in \acrshort{tst} or \acrshort{sws}.
     \item $R_{\text{Rules\_x}, \text{Outcome}}$: correlation between ML priority score \& percentage change in \acrshort{tst} or \acrshort{sws}.
    \item $\text{ML\_x}_i$: ML priority score (raw or adjusted) for the ith user.
    \item $\overline{\text{ML\_x}}$: mean of ML-priority scores (raw or adjusted) for all users.
    \item $\text{Rules\_x}_i$: rules-priority score (raw or adjusted) for the ith user.
    \item $\overline{\text{Rules\_x}}$: mean of rules-priority scores (raw or adjusted) for all users.
    \item $\text{Outcome}_i$:  $i$-th user \acrshort{tst} or \acrshort{sws} \% change (last 7 days vs baseline).
    \item $\overline{\text{Outcome}}$: mean \acrshort{tst} or \acrshort{sws} \% change (last 7 days vs baseline) for all.
    \item $n$: total number of users.
\end{itemize}
\end{mybox}

\begin{figure}[!t]
     \centering
     \includegraphics[width=\textwidth, keepaspectratio]{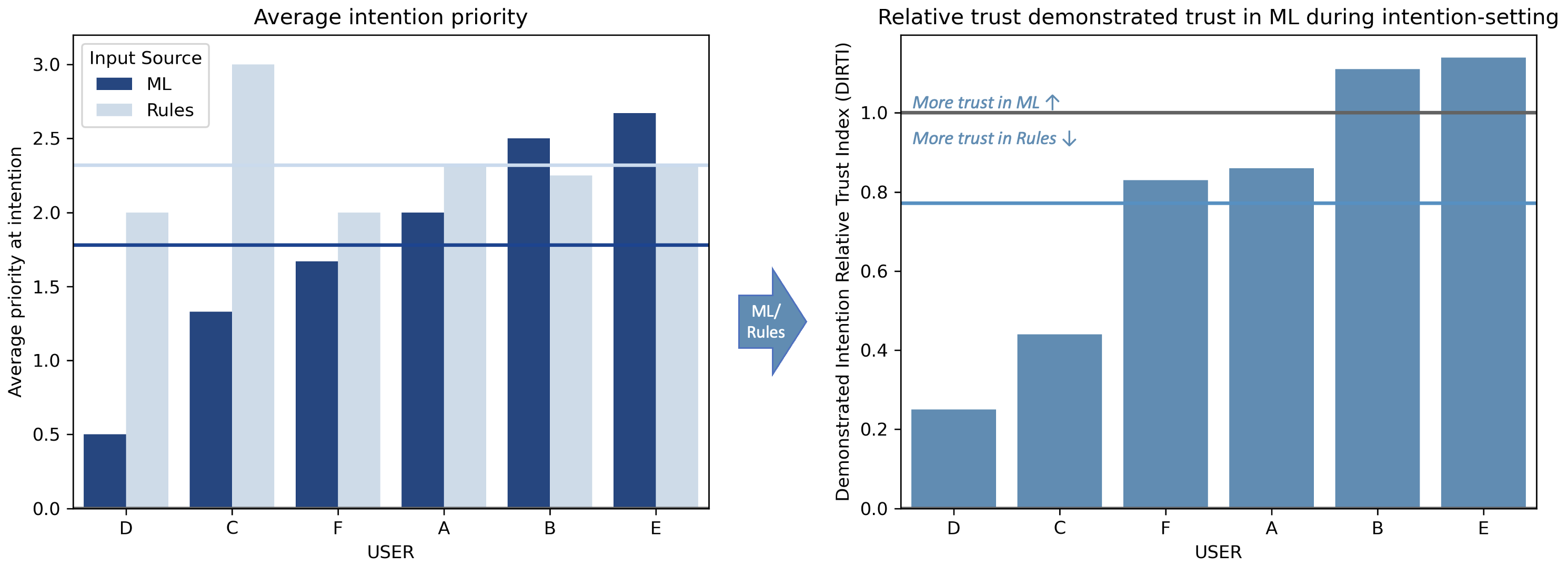} 
     
     \caption{\acrshort{dirti}: average \acrshort{ml} intervention priority (dark blue) divided by average rule priority (light blue) yields the relative trust demonstrated in \acrshort{ml} at intention on right. Blue lines are averages. 
}
\label{DIRTI}

\end{figure}

\acrfull{dirti} measures how users demonstrate trust in \acrshort{ml} through their intentions. 
Using two study participants on the left-hand side of Figure \ref{DIRTI} as an example, D chose 1 action with an \acrshort{ml} score of above zero on their \acrshort{lofm} horizontal axis; all other choices only scored highly on the rules axis. D's average rules score was 2 (pale blue) and \acrshort{ml} 0.5 (dark blue). Most of B's choices scored high on the \acrshort{ml} axis; some scored lower on rules, so they averaged 2.5 on \acrshort{ml} and 2.25 on rules. B placed more relative trust in \acrshort{ml} than D: B's \acrshort{dirti}, calculated by dividing their \acrshort{ml} score by their rules score, is 1.1; D's 0.25, as shown on the right-hand side of Figure \ref{DIRTI}. Although all participants could have chosen actions with at least equal \acrshort{ml} and rules scores, most relied more on rules than on \acrshort{ml} when they set intentions; only 2 had \acrshort{dirti} >1.

\begin{figure}[!p]
     \centering
     \includegraphics[width=1\textwidth, keepaspectratio]{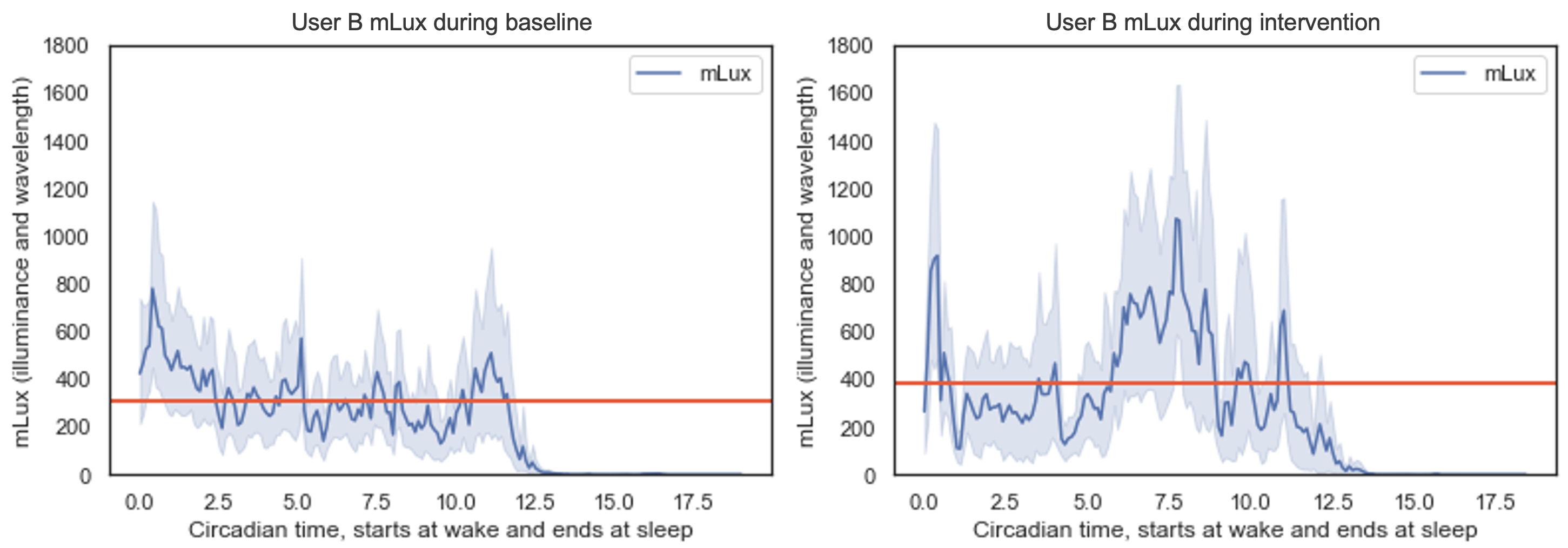} 
     
     \caption{B's mean light exposure from wake (CT0) to bedtime during baseline (left-hand side) and intervention (right-hand side), in blue (shaded area is 95th percentile). Red line is period mean.
}
\label{user_b_light}

\end{figure}

\begin{figure}[!p]
     \centering
     \includegraphics[width=1\textwidth, keepaspectratio]{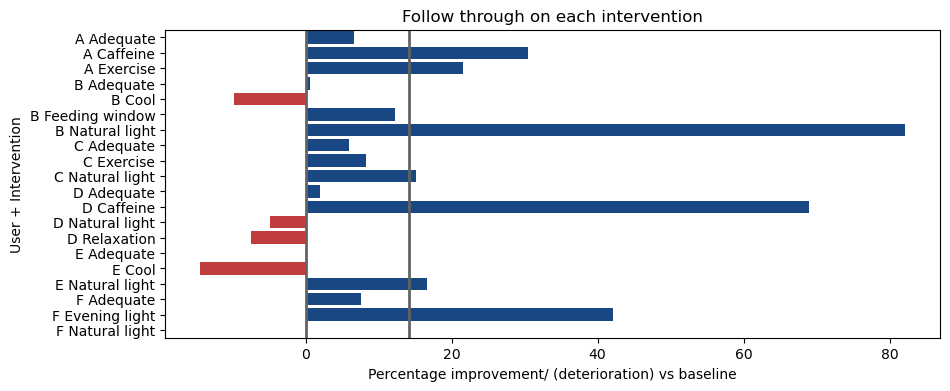} 
     
     \caption{Indexed percentage improvement vs baseline in blue; deterioration vs baseline in red.}
\label{compliance}

 \end{figure}

Demonstrating trust requires action, not just intention, which is what \acrfull{dafti} measures.
Some interventions required major changes to participants' daily routines. One commented `I don't need a sleep coach; I need a life coach' when planning how to follow through on their intentions. For example, as illustrated in Figure~\ref{user_b_light}: B made a big increase in their daily exposure in intervention compared to baseline, in red, and shifted a lot of it into the afternoon. 
Figure~\ref{compliance} shows participant follow-through, comparing the last 7 days of intervention to baseline. 
 The last 7 days was used because most participants had to work towards target behaviour in stages over the 4 weeks. The proliferation of blue indicates most showed some follow-through. D was very successful on caffeine; B on natural light. Some, in red, failed to follow through: D on 2; B's and E's bedroom temperatures increased, but they made other changes to cool their bodies down.  

\begin{figure}[!p]
     \centering
     \includegraphics[width=1\textwidth, keepaspectratio]{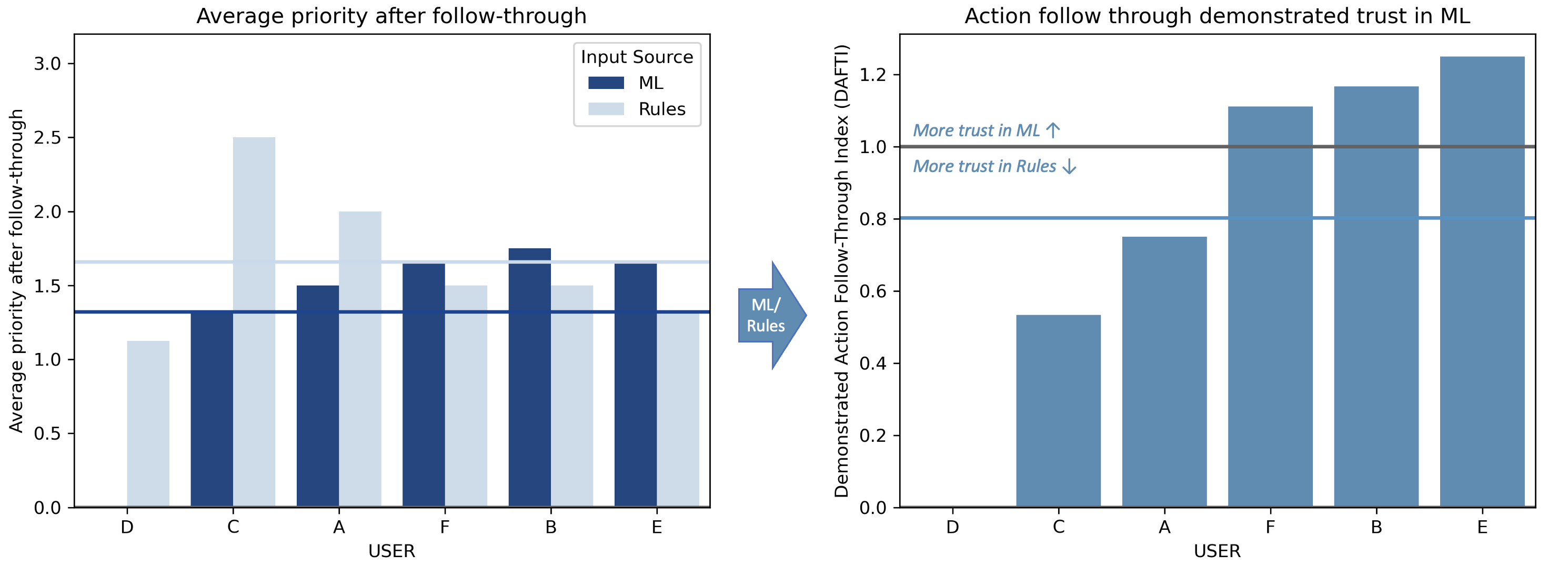} 
     
     \caption{\acrshort{dafti}: average \acrshort{ml} priority after follow-through (dark blue) divided by average rule priority (light blue) yields the relative trust demonstrated in \acrshort{ml} after follow-through (mid-blue).  Horizontal lines are averages. 
}
\label{DAFTI}

\end{figure}
 
The \acrshort{dafti} metric reflects action follow-through by discarding an intervention score if a participant fails to improve on baseline. 
For example, D acted on their caffeine intention, but failed to implement a relaxation routine, their only intervention with a positive \acrshort{ml} score. 
As shown in Figure~\ref{DAFTI}, D's average \acrshort{ml} score therefore dropped to 0, and their rules score dropped to 1.1 (light blue) because they also failed to follow through on their natural-light intention. Their relative demonstrated trust in \acrshort{ml} (\acrshort{dafti}), dividing their \acrshort{ml} score by their rules score, was 0. B followed through on all but one action, and their relative trust increased to 1.2. Three participants now showed more trust in \acrshort{ml} than rules (\acrshort{dafti}>1): F did a great job on evening light, a high-scoring ML intervention, and not so well on re-calibrating natural light levels, a high-scoring rules intervention.

Our \acrshort{ml} models were designed to be trustworthy, so the outstanding question is whether there is a relationship between the models' prioritisation of recommendations and the outcomes delivered, i.e. did participants who implemented high-scoring interventions get better outcomes? Was the risk worth it? 
The \acrfull{doti} measures whether the trust was worth it, reflecting the constraints faced once the participant acted. There will be a \textit{placebo effect} at play, which can be considerable, but there is no reason to assume it would not align to demonstrated trust.

\begin{figure}[!t]
     \centering
     \includegraphics[width=\textwidth, keepaspectratio]{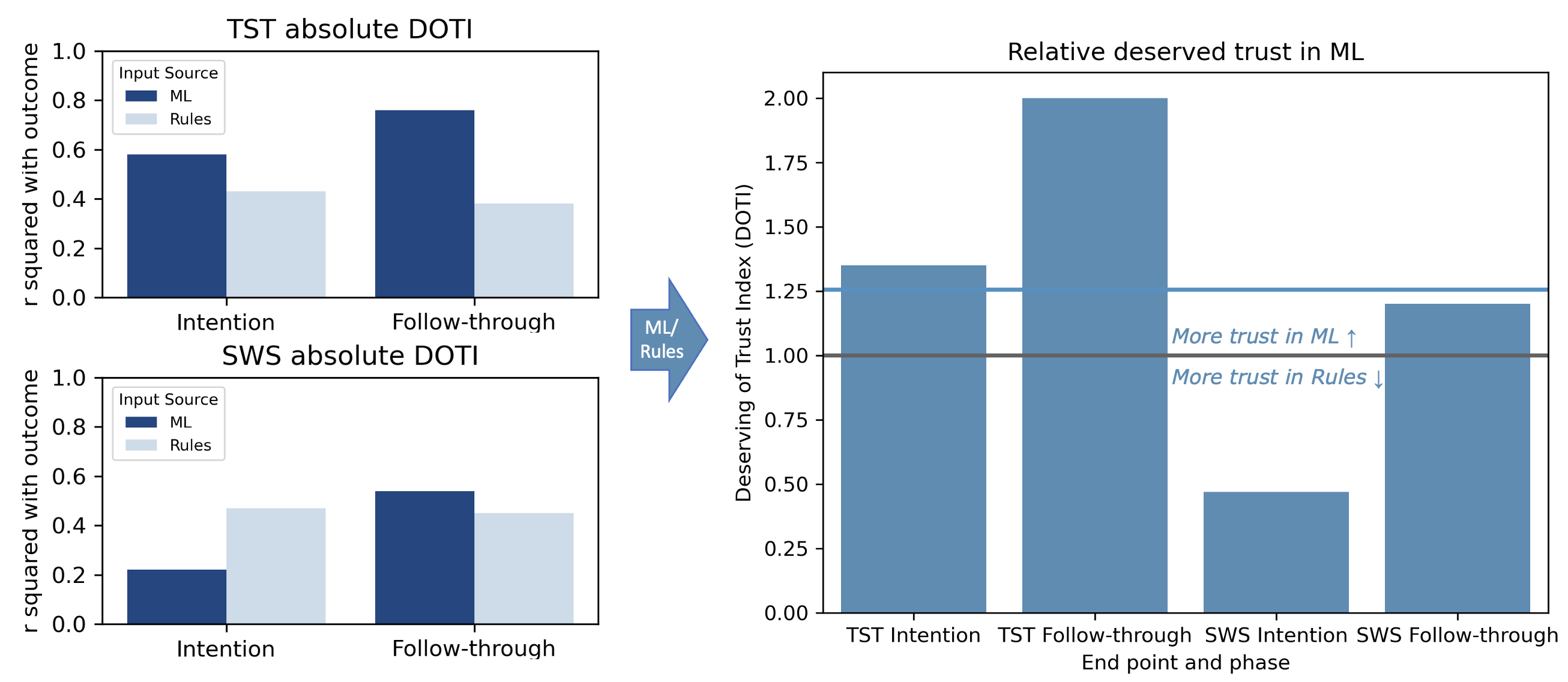} 
     
     \caption{ Absolute \acrshort{doti} on the left: \acrshort{tst} at top and \acrshort{sws} on bottom: correlation (${r}^2$) between intention and follow-through scores, and \% improvement in last 7 days' sleep end points relative to baseline. Relative \acrshort{doti} is on the right (\acrshort{ml} metric divided by rules-based). 
}
\label{DOTI}

\end{figure}

Demonstrated trust is shown on an absolute basis on the left of Figure~\ref{DOTI} – i.e. the correlation (${r}^2$) between each individual's average ML or rules score and their percentage increase in \acrshort{tst} and \acrshort{sws}. We compare baseline to the last seven days of intervention because most participants worked towards target behaviour in stages over four weeks. 
In this pilot, the strongest relationship was for \acrshort{ml}-interventions at follow-through (dark blue): \acrshort{tst} ${r}^2$ increases from 0.58 at intention to 0.76; \acrshort{sws} from 0.22 to 0.54. An increase is consistent with a trustworthy model because recommendations only work if people act on them.
Relative \acrshort{doti}, the ratio of the absolute \acrshort{ml} score and the rules score, is shown on the right. Trust was more deserved in the \acrshort{ml} algorithm for \acrshort{tst} than for \acrshort{sws}, which may have been influenced by lower \acrshort{sws} model accuracy. 
At follow-through, our \acrshort{ml} model deserved more trust than the rules-based one; had only intention been considered, the \acrshort{sws} rules-based model would have (incorrectly) been seen to deserve more trust. 

\section{Discussion}
\label{discussion}

The proposed architecture allows the \acrshort{ml} leap of faith to be measured and managed, contributing guidance on how to: 
\begin{itemize}[nosep]
    \item Build user intrinsic trust in \acrshort{ai} agents, by pre-validating data in an accessible format, and deciding their objectives. This functional compartmentalisation isolates algorithmic reasoning. 
    \item Build expert intrinsic trust, by providing a rules-based model that they test a priori against their mental model. If this were collectively-developed, it could become a reference agent. Industry or regulatory bodies could maintain the agent as critical infrastructure – e.g. for assurance audits or to train up new experts.
    Many domains already use rules-based models (e.g. banking), which could become reference standards.
    \item Bring the \acrshort{ml} leap of faith into sharp focus: a \acrshort{lofm} matrix visually and accessibly contrasts the output of the reference agent with the \acrshort{ml} model's a posteriori output, using the same data and objective function.
    This provides real transparency to system users: they should be inclined to trust where agents agree, and understand a leap of faith is only required where they disagree. 
    Experts can stress test the areas of disagreement, updating the rules-based agent with new insights, thus narrowing the leap of faith. Different \acrshort{ml} algorithms will require different leaps of faith, because a priori levels of expert insight differ, as do model accuracies.
    \item Assess demonstrated and deserved trust, by considering recommendation actions and outcomes. If trust is deserved, so relative \acrshort{doti} increases over time; \acrshort{dirti} and \acrshort{dafti} should also increase. These simple metrics allow trust comparisons across models, and flag the need for investigation if they trend down.  Industry regulators could track companies' \acrshort{ml}-model-trust metrics, especially \acrshort{doti}, to understand trends.  

\end{itemize}
Limitations include: 
\begin{itemize}[nosep]
    \item The matrix and metrics may be perceived as \textit{insufficiently technical}. Rather, they are accessible by design, so they can be used and understood by experts, businesses, and regulators as well as computer/ data scientists. They need to be applied at scale to reach empirical conclusions about trust, and what drives it, objectively.
    \item \textit{The proposed metrics do not measure attitudinal trust or model trustworthiness}. Instead they measure how these are reflected in the drivers of \acrshort{ml} adoption: decisions, actions, and outcomes. 
    If \acrshort{gdpr} (or similar) is not followed, trust metrics could be inappropriately sold and/ or used to discriminate based on people's trust inclination.
    \item \textit{Participant time for data validation} can be substantial, especially if data does not readily align with perception.
Our approach may be most suitable for professionals who use agent recommendations often, so their time commitment pays off as they focus on where there is a leap of faith to be overcome. The \textit{data-validation approach is not universally accessible}: our participants had at least one university degree and understood complex, novel concepts and images. Work is needed to evaluate accessibility for other levels of educational attainment.
    \item The \textit{time required to develop a reference agent} from scratch can be considerable, requiring: substantial programmer subject-matter education; individuals who can translate expert knowledge into rules; and time from experts on each iteration, and to test the resulting agent for different use cases.  
    \item Prioritisation, unlike direct regression/ classification comparisons, requires calibration if it is to be included in a \acrshort{lofm} matrix: \textit{different approaches should be tested}. 
    
\end{itemize}
\paragraph{Conclusion} The proposed architecture is a more practical, rigorous way of building trust in an \acrshort{ml} agent than trying to explain it to people who have a different (probably unarticulated) mental model. Critical novel architecture features include: user data validation and objective-function specification; an explicit reference standard agent; a \acrshort{lofm} matrix; and demonstrated- and deserved-trust metrics.

\newpage

\bibliography{main}

\newpage

\newpage

 \section{Abbreviations}

\setlength{\glsdescwidth}{0.7\textwidth}
\setglossarystyle{long}
\setlength\LTleft{0pt}
\setlength\LTright{0pt}
\setlength\glsdescwidth{0.8\hsize}
\printglossary[type=\acronymtype, title = ]


\end{document}